\documentclass[onecolumn,showpacs,preprintnumbers,amsmath,amssymb,showpacs,showcolor]{revtex4}

\usepackage{graphicx}
\usepackage{dcolumn}
\usepackage{bm}
\usepackage{amssymb}
\usepackage{color}

 \def\tskip{\setlength{\tskip}{5pt}}
\def\colwidth{\setlength{\colwidth}{3.5in}}

\def\prd{Phys. Rev. D}
\def\pr{Phys. Rep.~}
\def\prl{Phys. Rev. Lett.~}
\def\plb{Phys. Lett. B}

\def\apj{Astrophys. J.~}

\def\apjs{Astrophys. J. Suppl. Ser.~}

\def\cqg{Class. Quant. Grav. ~}
\def\lnp{Lect. Notes Phys.~}

\newcommand{\lsim}{\mathrel{\hbox{\rlap{\lower.55ex\hbox{$\sim$}} \kern-.3em \raise.4ex \hbox{$<$}}}}
\newcommand{\gsim}{\mathrel{\hbox{\rlap{\lower.55ex\hbox{$\sim$}} \kern-.3em \raise.4ex \hbox{$>$}}}}
\newcommand{\beq}{\begin{equation}}
\newcommand{\eeq}{\end{equation}}
\newcommand{\be}{\begin{equation}}
\newcommand{\ee}{\end{equation}}
\newcommand{\bes}{\begin{equation*}}
\newcommand{\ees}{\end{equation*}}
\newcommand{\beqa}{\begin{eqnarray}}
\newcommand{\eeqa}{\end{eqnarray}}
\newcommand{\bea}{\begin{eqnarray}}
\newcommand{\ena}{\end{eqnarray}}

\newcommand{\mpl}{m_\mathrm{Pl}}

\def\mpl{m_{\mathrm{Pl}}}

\begin{document}

\title{Constraint on the early Universe by relic gravitational waves: From pulsar timing observations}

\author{Wen Zhao}
\email{wzhao7@mail.ustc.edu.cn} \affiliation{Department of
Physics, Zhejiang University of Technology, Hangzhou 310014,
P.R.China}\affiliation{Niels Bohr Institute, Copenhagen
University, Blegdamsvej 17, Copenhagen DK-2100, Denmark}


\begin{abstract}
Recent pulsar timing observations by the Parkers Pulsar Timing
Array and European Pulsar Timing Array teams obtained the
constraint on the relic gravitational waves at the frequency
$f_*=1/{\rm yr}$, which provides the opportunity  to constrain
$H_*$, the Hubble parameter when these waves crossed the horizon
during inflation. In this paper, we investigate this constraint by
considering the general scenario for the early Universe: we assume
that the effective (average) equation-of-state $w$ before the big
bang nucleosynthesis stage is a free parameter. In the standard
hot big-bang scenario with $w=1/3$, we find that the current PPTA
result follows a bound $H_*\leq 1.15\times10^{-1}\mpl$, and the
EPTA result follows $H_*\leq 6.92\times10^{-2}\mpl$. We also find
that these bounds become much tighter in the nonstandard scenarios
with $w>1/3$. When $w=1$, the bounds become
$H_*\leq5.89\times10^{-3}\mpl$ for the current PPTA and
$H_*\leq3.39\times10^{-3}\mpl$ for the current EPTA. In contrast,
in the nonstandard scenario with $w=0$, the bound becomes
$H_*\leq7.76\mpl$ for the current PPTA.
\end{abstract}



\maketitle


\section{Introduction \label{section1}}

A stochastic background of relic gravitational waves, generated
during the early inflationary stage, is a necessity dictated by
general relativity and quantum mechanics
\cite{grishchuk1974,starobinsky1979,mukhanov1992}. The relic
gravitational waves have a wide range spreading spectra, and their
amplitudes depend only on the Hubble parameter in the inflationary
stage, when the waves crossed the horizon, and the expansion
history of Universe after the waves reentered the horizon. So
their detection provides a direct way to study the physics in the
early Universe in both stages, during and after the inflation.

Recently, there have been several experimental efforts to
constrain the amplitude of relic gravitational waves in the
different frequencies. The current observations of cosmic
microwave background (CMB) radiation by the WMAP satellite place
an interesting bound on the so-called tensor-to-scalar ratio $r\le
0.20$ \cite{komatsu2011,zhaogrishchuk2010}, which is equivalent to
the constraint on the energy density $\Omega_{\rm gw}(f)$ of relic
gravitational waves at the lowest frequency range $f \sim
10^{-17}$Hz. Among various direct observations, LIGO S5 has also
experimentally obtained so far the most stringent bound
$\Omega_{\rm gw}(f)\le 6.9\times10^{-6}$ around $f\sim 100$Hz
\cite{ligo2009}. In addition, there are two bounds on the
integration $\int \Omega_{\rm gw}(f)d\ln f \lsim1.5\times
10^{-5}$, obtained by the Big Bang nucleosynthesis (BBN)
observation \cite{bbn1999} and the CMB observation \cite{cmb2006}.
These bounds have been used to constrain the Hubble parameter (or
the potential density of inflaton) in the inflationary stage, when
the corresponding waves crossed the horizon
\cite{komatsu2011,ligo2009,zhang2010}.

The timing studies on the millisecond pulsars provide a unique way
to constrain the amplitude of gravitational waves in the frequency
range $f\in (10^{-9}, 10^{-7})$Hz \cite{timing}. Recently, the
Parkers Pulsar Timing Array (PPTA) team and the European Pulsar
Timing Array (EPTA) team have reported their observational results
on the stochastic background of gravitational waves and given the
upper limit of $\Omega_{\rm gw}(f)$ at the frequency $f=1/{\rm
yr}$ \cite{ppta,epta}. In this paper, we shall infer from these
bounds the constraint on $H$, the Hubble parameter at the waves'
horizon-crossing time during inflation. In the calculation, we
have considered a general early cosmological model, i.e., we
assume the effective (average) equation-of-state $w$ before the
BBN stage can be of any value, which includes a wide range of
cosmological scenarios. The derived bound of $H$ would limit
various inflation models.


\section{Relic gravitational waves in the standard hot big-bang universe \label{section2}}

Incorporating the perturbation to the spatially flat
Friedmann-Robertson-Walker (FRW) spacetime, the metric is
 \beq\label{metric}
 ds^2=a^{2}(\eta) \left[d\eta^2-(\delta_{ij}+h_{ij}dx^idx^j)\right],
 \eeq
where $a$ is the scale factor of the universe, and $\eta$ is the
conformal time, which relates to the cosmic time by $a d\eta=dt$.
The perturbation of spacetime $h_{ij}$ is a $3\times 3$ symmetric
matrix. The gravitational-wave field is the tensorial portion of
$h_{ij}$, which is transverse-traceless $\partial_ih^{ij}=0$,
$\delta^{ij}h_{ij}=0$.

{Relic gravitational waves satisfy the linearized evolution
equation \cite{grishchuk1974}:
 \beq\label{evolution}
 \partial_{\mu}(\sqrt{-g}\partial^{\mu}h_{ij})=-16\pi G \pi_{ij}.
 \eeq
The anisotropic portion $\pi_{ij}$ is the source term, which can
be given by the relativistic free-streaming gas
\cite{weinberg2003} and the scalar field in the preheating stage
\cite{preheating}. However, it has been deeply discussed that the
relativistic free-streaming gas can only affect the relic
gravitational waves at the frequency range
$f\in(10^{-16},~10^{-10})$Hz, which could be detected by the
future CMB observations \cite{zzx2009}. The generation of
stochastic background of gravitational waves in the preheating
stage has also been deeply studied (see, for instance,
\cite{preheating}), where the gravitational radiation was produced
in interactions of classical waves created by resonant decay of a
coherently oscillating field. However, it was found that the
typical frequencies of this kind of gravitational waves are quite
high, i.e. $f>10^{4}$Hz. Even if the model with low energy $H\sim
100$GeV is considered, the gravitational waves are important only
at the frequency range $f\sim 1$Hz \cite{preheating}, which could
be detected by the future laser interferometer detectors. So, both
effects cannot obviously influence the relic gravitational waves
at the frequency $f\in (10^{-9},~10^{-7})$Hz}. For these reasons,
in this paper we shall ignore the contribution of the external
sources. So the evolution of gravitational waves is only dependent
on the scale factor and its time derivative. It is convenient to
Fourier transform the equation as follows:
 \be\label{fourier}
 h_{ij}(\eta,\vec{x})=\int \frac{d^3 \vec{k}}{(2\pi)^{3/2}}\sum_{s=+,\times}
 \left[
 h_k(\eta) \epsilon^{(s)}_{ij}c^{(s)}_{\vec{k}}e^{i\vec{k}\cdot\vec{x}}+c.c.
 \right],
 \ee
where $c.c.$ stands for the complex conjugate term. The
polarization tensors are symmetry, transverse-traceless
$k^i\epsilon_{ij}^{(s)}(\vec{k})=0$,
$\delta^{ij}\epsilon_{ij}^{(s)}(\vec{k})=0$, and satisfy the
conditions
$\epsilon^{(s)ij}(\vec{k})\epsilon_{ij}^{(s')}(\vec{k})=2\delta_{ss'}$
and $\epsilon_{ij}^{(s)}(-\vec{k})=\epsilon_{ij}^{(s)}(\vec{k})$.
Since the relic gravitational waves we will consider are isotropy,
and each polarization state is the same, we have denoted
$h_{\vec{k}}^{(s)}(\eta)$ by $h_k(\eta)$, where $k=|\vec{k}|$ is
the wavenumber of the gravitational waves, which relates to the
frequency by $k\equiv 2\pi f$. (The present scale factor is set
$a_0=1$). So Eq. (\ref{evolution}) can be rewritten as
 \beq\label{h-evolution}
{h_k}''+2\frac{{a'}}{a}{h_k}'+k^2h_k=0,
 \eeq
where the {\it prime} indicates a conformal time derivative
$d/d\eta$. For a given wavenumber $k$ and a given time $\eta$, we
can define the transfer function $t_f$ as
 \be\label{tf-define}
 t_f(\eta,k)\equiv h_k(\eta)/h_k(\eta_i),
 \ee
where $\eta_i$ is the initial conformal time. This transfer
function can be obtained by solving the evolution equation
(\ref{h-evolution}).

The strength of the gravitational waves is characterized by the
gravitational-wave energy spectrum,
 \beqa
 \Omega_{\rm gw} \equiv \rho_{\rm gw}/\rho_0,
 \eeqa
where $\rho_{\rm gw}=\frac{1}{32\pi
G}\langle\dot{h}_{ij}\dot{h}^{ij}\rangle$, the critical density is
$\rho_0=\frac{3H_0^2}{8\pi G}$, and $H_0=100h\cdot{\rm km}\cdot
s^{-1}\cdot{\rm Mpc}^{-1}$ is the current Hubble constant. Using
Eqs. (\ref{fourier}) and (\ref{tf-define}), the energy density of
gravitational waves can be written as \cite{page2006}
 \be
 \rho_{\rm gw}=\int \frac{d k}{k} \frac{P_t(k)\dot{t}^2_f(\eta_0,k)}{32\pi G},
 \ee
where $P_t(k)\equiv \frac{2k^3}{\pi^2}|h_k(\eta_i)|^2$ is the
so-called primordial power spectrum of relic gravitational waves.
Thus, we derive that the current energy density of relic
gravitational waves
 \be\label{omega-gw}
 \Omega_{\rm gw} \equiv \int d\ln{k} ~\Omega_{\rm gw} (k),
{~\rm and~}
 \Omega_{\rm gw} (k) = \frac{P_t(k)}{12 H_0^2}\dot{t}_f^2(\eta_0,k)
 \ee
where the  {\it dot} indicates a cosmic time derivative $d/dt$.

Now, let us discuss the terms $P_t(k)$ and $t_f(\eta_0,k)$
separately. The primordial power spectrum of relic gravitational
waves is usually assumed to be power-law as follows:
 \be\label{pt1}
 P_t(k) = A_t(k_*)\left(\frac{k}{k_*}\right)^{n_t}.
 \ee
This is a generic prediction of a wide range of scenarios of the
early Universe, including the inflation models. Here, we should
mention that there might be deviations from power-law if we
consider the relic gravitational waves in a fairly large wave
number span. \emph{In this paper, as a conservative consideration,
we assume this form is held only when $k$ is very close to the
pivot wavenumber $k_*$}. In the above expression, $n_t$ is the
spectral index when $k\rightarrow k_*$. ($n_t=0$ corresponds to
the scale-invariant power spectrum.) $A_t(k_*)$ is directly
related the value of the Hubble parameter $H$ at time when
wavelengths corresponding to the wavenumber $k_*$ crossed the
horizon \cite{grishchuk1974,mukhanov1992,peiris2003},
 \be\label{pt2}
 A^{1/2}_t(k_*)=\left.\frac{4}{\sqrt{\pi}}\frac{H_*}{\mpl}\right|_{k_*=a_*H_*},
 \ee
where $\mpl\equiv1/\sqrt{G}$ is the Planck mass.

Now, let us turn to the transfer function $t_f$, defined in
(\ref{tf-define}), which describes the evolution of gravitational
waves in the expanding Universe. From Eq. (\ref{h-evolution}), we
find that this transfer function can be directly derived, so long
as the scale factor as a function of time is given. Actually, the
analytical or numerical forms of $t_f$ have been discussed by a
number of authors (see, for instance,
\cite{grishchuk2000,zhang2005,tong2009,watanabe2006}).

In this paper, we shall use the following analytical approximation
for this transfer function. It has been known that, during the
expansion of the Universe, the mode function $h_k(\eta)$ of the
gravitational waves behaves differently in two regions
\cite{grishchuk2000}. When waves are far outside the horizon, i.e.
$k\ll aH$, the amplitude of $h_k$ keeps constant, and when inside
the horizon, i.e. $k\gg aH$, the amplitude is damping with the
expansion of Universe, i.e., $h_k\propto 1/a(\eta)$. In the
standard hot big-bang cosmological model, we assume that the
inflationary stage is followed by a radiation dominant stage, and
then the matter dominant stage and the $\Lambda$ dominant stage.
In this scenario, by numerically integrating Eq.
(\ref{h-evolution}), one finds that the damping function
$\dot{t}_f$ can be approximately described by the following form
\cite{turner1993,zhao2006,efstathiou2006,giovannini2009}
 \begin{equation}\label{tf2}
 \dot{t}_f(\eta_0,k)=\frac{-3j_2(k\eta_0)}{k\eta_0}\frac{\Omega_m}{\Omega_{\Lambda}}
 \sqrt{1+1.36(\frac{k}{k_{eq}})+2.50(\frac{k}{k_{eq}})^2},
 \end{equation}
where $k_{eq}=0.073\Omega_m h^2{\rm Mpc}^{-1}$ is the wavenumber
corresponding the Hubble radius at the time that matter and
radiation have equal energy density, and $\eta_0=1.41\times
10^{4}{\rm Mpc}$ is the present conformal time. The factor
$\Omega_m/\Omega_{\Lambda}$ encodes the damping effect due to the
recent accelerating expansion of the Universe
\cite{zhang2005,zhao2006}. In this damping factor, we have ignored
the small effects of neutrino free-streaming \cite{weinberg2003}
and various phase transition \cite{watanabe2006}.

We can define a new function
 \be\label{macT}
 \mathcal{T}(k) \equiv \dot{t}_f(\eta_0,k)/\sqrt{12H_0^2}.
 \ee
So, the current density of relic gravitational waves becomes
$\Omega_{\rm gw}(k)=P_t(k)\mathcal{T}^2(k)$. In this paper, we
shall focus on the wavenumber $k\gg k_{eq}$. In this range, we
have \cite{zhao2006,efstathiou2006,giovannini2009}\footnote{ In
our previous work \cite{zhao2006}, only the amplitudes of the
quick oscillating gravitational waves are considered. However,
here we have considered the average energy density of
gravitational waves. The difference between them is a factor $2$.}
 \be\label{tf3}
 \mathcal{T}^2(k)=\frac{15}{16}
 \left(\frac{\Omega_m}{\Omega_{\Lambda}}\right)^2
 \frac{1}{H_0^2 \eta_0^4 k_{eq}^2}.
 \ee
and a parameterized form for the current density of relic
gravitational waves
 \be\label{e71}
 \Omega_{\rm gw}(k) = \left(\frac{H_*}{\mpl}\right)^2
 \left(\frac{k}{k_*}\right)^{n_t}
 \left(\frac{15}{\pi}\frac{1}{H_0^2\eta_0^4
 k_{eq}^2}\frac{\Omega_m^2}{\Omega_{\Lambda}^2}\right).
 \ee

For the wavenumber $k=k_*$, the value of $\Omega_{\rm gw}(k_*)$
depends only on the value of $H_*$. So, in this standard scenario,
an observational bound on the $\Omega_{\rm gw}(k_*)$ corresponds
to a bound on the Hubble parameter $H_*$, which will be shown
clearly in Sec. \ref{section4}.

\section{Damping factor in the general model of the early universe\label{section3}}

Although, in the standard hot big-bang universe, a radiation
dominant stage is always assumed after the inflationary stage,
there is no observational evidence to show this is held before the
BBN stage. Actually, this assumption can be violated in a number
of cases, for example, the existence of the reheating stage
\cite{grishchuk2000}, or the existence of the cosmic phase
transition \cite{watanabe2006}. So, in general, before the BBN
stage, one can assume that the average equation-of-state of the
Universe is $w$, and the scale factor satisfies a simple power-law
form
 \be\label{genel-a}
 a\propto \eta^{1+\beta}.
 \ee
The constant $\beta$ relates to $w$ by $\beta=(-3w+1)/(3w+1)$.
Obviously, when $w=1/3$, i.e. $\beta=0$, it returns to the
standard model. However, if the Universe is dominated by the
kinetic energy of inflaton, one has $w=1$ and $\beta=-1/2$. On the
other hand, for a matter dominated era, one has $w=0$ and
$\beta=1$.

Now, let us discuss the evolution of relic gravitational waves in
this general cosmological model. In principle, it can be done by
directly solving Eq. (\ref{h-evolution}). In this paper, in order
to avoid the complicated numerical calculation, we give an
approximate method as below.

We consider the wave $h_k$ with the wavenumber $k$, which crossed
the horizon at $a=a_k$ and the corresponding Hubble parameter is
$H_k$. So one has $k=a_kH_k/a_0$. One knows that, when the waves
are in the horizon, $h_k\propto 1/a(\eta)$, damping with the
expansion of the Universe, and when the waves are out the horizon,
$h_k=constant$, keeping its initial value. So one can define a
ratio, which accounts for the damping of the gravitational waves,
 \be
 \frac{h_k(\eta_0)}{h_k(\eta_i)} = \frac{a_k}{a_0} =
 \frac{a_k}{a_b} \frac{a_b}{a_0},
 \ee
where $a_b$ is the scale factor at the temperature of Universe
being $1{\rm MeV}$, i.e. the BBN stage.

In the standard model, where $\beta=0$ in (\ref{genel-a}) is
assumed (i.e. $w=1/3$, the radiation dominant stage), we have
 \be\label{e81}
 \frac{H_k}{H_b}=\left(\frac{a_b}{a_k}\right)^2,
 \ee
where $H_b$ is the Hubble parameter in the BBN stage. Taking into
account the relation $k=a_kH_k/a_0$, we obtain that
 \be\label{e82}
 \frac{h_k(\eta_0)}{h_{k}(\eta_i)}=\frac{a_b}{a_0}\left(\frac{a_bH_b}{a_0k}\right).
 \ee

However, in the general case with $\beta\neq 0$, we assume $h_k$
crossed the horizon at $a=\tilde{a}_k$ and the corresponding
Hubble parameter is $\tilde{H}_k$. (Note that, in general
$\tilde{a}_k\neq a_k$ and $\tilde{H}_k\neq H_k$, but
$k=\tilde{a}_k\tilde{H}_k/a_0$ is still satisfied.) From the
equation in (\ref{genel-a}), it follows that
 \[
 \frac{\tilde{H}_k}{H_b}=\left(\frac{a_b}{\tilde{a}_k}\right)^{\frac{2+\beta}{1+\beta}}.
 \]
So, in this general case, we have
 \be\label{e92}
 \frac{h_k(\eta_0)}{h_{k}(\eta_i)}=\frac{\tilde{a}_k}{a_0}=\frac{a_b}{a_0}\left(\frac{a_bH_b}{a_0k}\right)^{1+\beta}.
 \ee

Comparing Eqs. (\ref{e92}) and (\ref{e82}), we can define the
damping faction $D(k)$ as follows
 \beqa
 D(k)&\equiv&\left(\frac{h_k(\eta_0)}{h_{k}(\eta_i)}\right)_{general}/\left(\frac{h_k(\eta_0)}{h_{k}(\eta_i)}\right)_{standard}\\
 &=&\left(\frac{a_bH_b}{a_0
 H_0}\right)^{\beta}\left(\frac{H_0}{k_*}\right)^{\beta}\left(\frac{k}{k_*}\right)^{-\beta}\label{e01}
 \eeqa

Thus, in this general scenario, the current density of relic
gravitational waves becomes
 \be\label{e11}
 \Omega_{\rm gw}(k)=P_t(k)\mathcal{T}^2(k) D^2(k),
 \ee
which satisfies $\Omega_{\rm gw}(k)\propto k^{n_t-2\beta}$ when
$k$ is close to $k_*$. Using the formulae in (\ref{pt1}),
(\ref{pt2}), (\ref{tf3}), (\ref{e01}), and substituting the
cosmological parameters ($h=0.702$, $T_{\rm CMB}=0.276$K,
$\Omega_{\Lambda}=0.725$, $\Omega_{m}=0.275$, and $z_{eq}=3454$
)\cite{komatsu2011}, we get the following simple result
 \be\label{e13}
 \log_{10}\Omega_{\rm gw}(k_*) = 1.25 -\frac{13.48}{3w+1} +
 2\log_{10}(\frac{H_*}{\mpl}),
 \ee
where $k_*\equiv 2\pi f_*$, and $f_*=1/{\rm yr}$ is used. In Sec.
\ref{section4}, we shall compare this with the observational
results.

\section{Constraint by the pulsar timing observations\label{section4}}

\begin{figure}[t]
\centerline{\includegraphics[width=15cm,height=12cm]{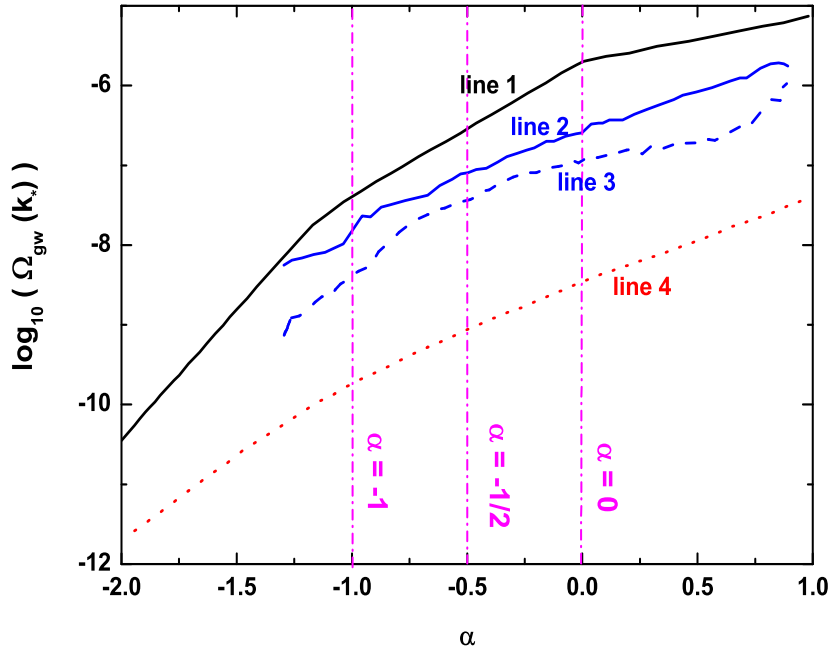}}
\caption{The upper limit of $\Omega_{\rm gw}(k_*)$ as a function
of the parameter $\alpha$. The black solid line (i.e. line 1) is
for current PPTA 2$\sigma$ result \cite{ppta}, the blue solid line
(i.e. line 2) is for current EPTA 2$\sigma$ result \cite{epta},
the blue dashed line (i.e. line 3) is for current EPTA 1$\sigma$
result \cite{epta}, and the red dotted line (i.e. line 4) is for
future PPTA 2$\sigma$ result \cite{ppta}. }\label{figure2}
\end{figure}

Pulsar timing observations provide a unique opportunity to study
the gravitational waves at the frequency range $f\in
(10^{-9},~10^{-7}){\rm Hz}$. In 2006, Jenet et al. have analyzed
the PPTA data and archival Arecibo data for several millisecond
pulsars. By focusing on the gravitational waves with the wavenumer
$k_*$ (where $k_*=2\pi f_{*}$ and $f_*=1/{\rm yr}$), and assuming
the density of gravitational waves satisfies $\Omega_{\rm
gw}(k)=k^{2+2\alpha}$ at around $k\sim k_*$, the authors obtained
the 2$\sigma$ upper limit on $\Omega_{\rm gw}(k_*)$ as a function
of $\alpha$ \cite{ppta}, which has been shown in Fig.
\ref{figure2} (black solid line). This figure shows that
$\Omega_{\rm gw}(k_*)\le 4.05\times 10^{-8}$ when $\alpha=-1$.
However, this upper bound increases to be $1.98\times 10^{-6}$
when $\alpha=0$.

Recently, this upper limit has been updated. In \cite{epta}, the
authors have used the current data from the EPTA to determine an
upper limit on the stochastic gravitational-wave background as a
function of the spectral slope $\alpha$. The 1$\sigma$ and
2$\sigma$ bounds are shown in Fig. \ref{figure2} (blue lines),
which are slightly lower than those in PPTA case for any given
$\alpha$.

It is interesting that in \cite{ppta}, the authors have also
investigated the possible upper limit (or a definitive detection)
of stochastic background of gravitational waves by using the
potential completed PPTA data-sets (20 pulsars with an rms timing
residual of 100 ns over 5 years). We have also plotted this
potential upper limit in Fig. \ref{figure2} (red dotted line).

Now, let us compare these observations with the analytical
formulae of relic gravitational waves in Sec. \ref{section3}.
Firstly, it is necessary to relate the parameter $\alpha$ with the
theoretical models. In Sec. \ref{section3}, Eq. (\ref{e11}) shows
that $\Omega_{\rm gw}(k)\propto k^{n_t-2\beta}$, where
$\beta=(-3w+1)/(3w+1)$. Comparing this with the assumed form
$\Omega_{\rm gw}(k)\propto k^{2+2\alpha}$, we get the interesting
relation
 \be\label{e22}
 \alpha=\frac{n_t}{2}-\frac{2}{3w+1}.
 \ee
This relation shows that, in the standard hot big-bang scenario
with $w=1/3$, and the scale-invariant primordial power spectrum
with $n_t=0$, we have $\alpha=-1$. In this case, let us use the
bounds of gravitational waves to constrain the Hubble parameter
$H_*$ in the inflationary stage. Taking into account the formula
in Eq. (\ref{e13}) and using $w=1/3$, we obtain the 2$\sigma$
upper limit of $H_*$, i.e. $H_*\le1.15\times 10^{-1}\mpl$ for the
current PPTA case, $H_*\le6.92\times 10^{-2}\mpl$ for the current
EPTA case, and the future PPTA is expected to give
$H_*\le7.94\times 10^{-3}\mpl$. These results are listed in Table
\ref{table2}.

\begin{figure}[t]
\centerline{\includegraphics[width=15cm,height=12cm]{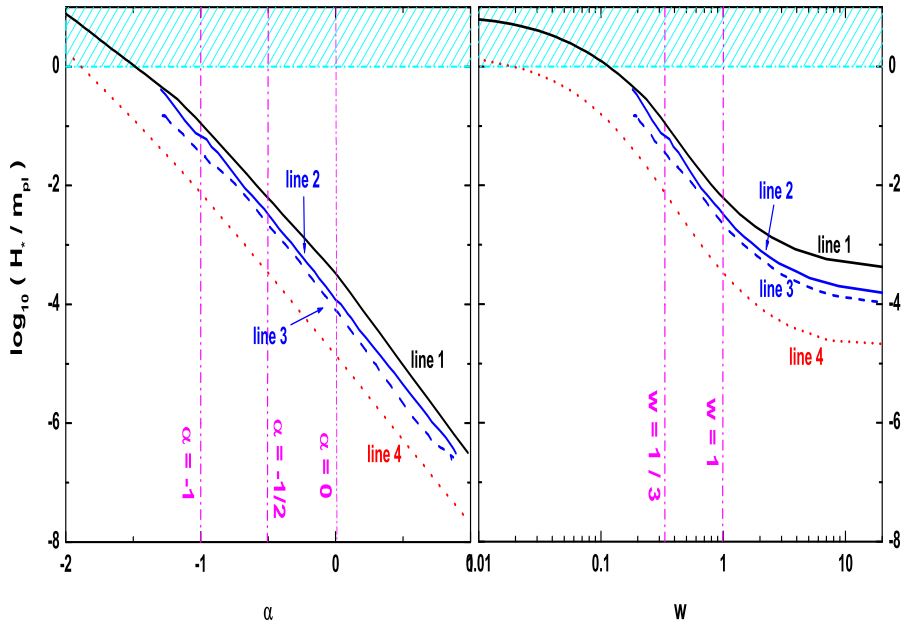}}
\caption{The upper limit of the Hubble parameter $H_*$ as a
function of the parameter $\alpha$ (left panel) and the average
equation-of-state $w$ (right panel), where we have assumed
$n_t=0$. In both panels, the black solid lines (i.e. line 1) are
for current PPTA 2$\sigma$ result, the blue solid lines (i.e. line
2) are for current EPTA 2$\sigma$ result, the blue dashed lines
(i.e. line 3) are for current EPTA 1$\sigma$ result, and the red
dotted lines (i.e. line 4) are for future PPTA 2$\sigma$ result.
}\label{figure3}
\end{figure}

\begin{figure}[t]
\centerline{\includegraphics[width=15cm,height=12cm]{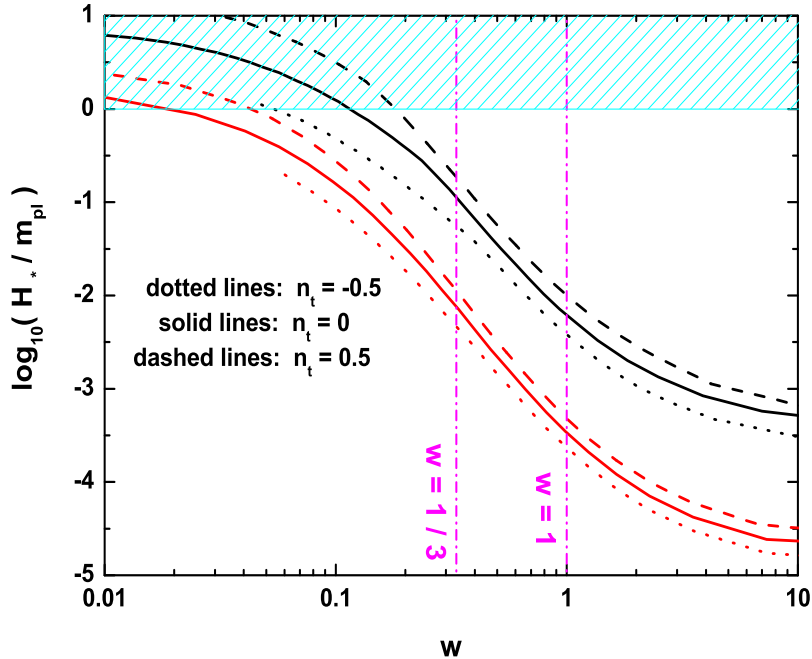}}
\caption{The upper limit of the Hubble parameter $H_*$ as a
function of the average equation-of-state $w$, where we have
considered three cases with $n_t=-0.5$, $0$, $0.5$. The black
lines are for current PPTA 2$\sigma$ result, and the red lines are
for future PPTA 2$\sigma$ result.}\label{figure4}
\end{figure}

Although, the inflation models always predict the nearly same
Hubble parameter throughout the inflationary stage, it is
necessary to constrain $H$, the Hubble parameter at quite
different stages of inflation, which encodes the evolution
information of inflaton. Here, let us compare the bound of $H$
inferred from pulsar timing with those obtained in CMB
observations and LIGO observations. The recent CMB observations by
the WMAP satellite provide the constraint on the tensor-to-scalar
ratio $r\leq 0.20$ \cite{komatsu2011}, which is equivalent to the
bound of $H/\mpl\leq6.92\times10^{-6}$, where $H$ is the Hubble
parameter of inflation when the waves with frequency
$f=1.94\times10^{-17}$Hz crossed the horizon. The recent LIGO 5S
reported so far the tightest constraint $\Omega_{\rm gw}(f)\leq
6.9\times 10^{-6}$ on relic gravitational waves at the frequency
$f\simeq 100$Hz \cite{ligo2009}, which corresponds to
$H/\mpl\leq1.46$. Comparing with these results, we find that the
current and the potential future pulsar timing constraints on $H$
are quite tighter than that of LIGO, but much looser than the CMB
constraint.

\begin{table}
\caption{The 2$\sigma$ upper limit of the quantity $H_*/\mpl$
inferred from various pulsar timing observations. In this table,
we have assumed $n_t=0$. }
\begin{center}
\label{table2}
\begin{tabular}{|c|c|c|c|}
    \hline
     & Current PPTA & Current EPTA & Future PPTA   \\
    \hline
   $w=0$ &   $7.76$   &  ......    &     $1.41$          \\
    \hline
   $w=1/3$ &   $1.15\times 10^{-1}$   &  $6.92\times 10^{-2}$    &     $7.94\times 10^{-3}$          \\
    \hline
    $w=1$ &     $5.89\times 10^{-3}$       &  $3.39\times 10^{-3}$            &      $3.55\times 10^{-4}$       \\
    \hline
   $w\rightarrow\infty$ &    $3.16\times 10^{-4}$       &  $1.20\times 10^{-4}$         &   $1.41\times 10^{-5}$          \\
   \hline
\end{tabular}
\end{center}
\end{table}

Now, let us relax the assumptions of the early Universe. We only
assume $n_t=0$, which is approximately held in a wide range of
inflation models. So, we can constrain the Hubble parameter $H_*$
in a wide range of $w$ by the following inequality
 \be\label{e23}
 \log_{10}(\frac{H_*}{\mpl}) \le
 \frac{1}{2}(U(\alpha)-6.74\alpha-1.25),
 \ee
where $U(\alpha)$ is the upper limit of $\Omega_{\rm gw}(k_*)$
based on the pulsar timing observations, which is a function of
the parameter $\alpha$ (in this case, $\alpha$ relates to $w$ by
the relation $\alpha=-2/(3w+1)$). The bounds of $H_*$ as functions
of $\alpha$ (left panel) and $w$ (right panel) are shown in Fig.
\ref{figure3}. These bounds in three special cases with $w=0$
(i.e. $\alpha=-2$), $w=1$ (i.e. $\alpha=-1/2$) and
$w\rightarrow\infty$ (i.e. $\alpha=0$) are also listed in Table
\ref{table2}. Clearly, we find that a larger $w$ corresponds to a
tighter bound of $H_*$. Especially, in the limit case with
$w\rightarrow\infty$, the current EPTA gives the constraint
$H_*\leq1.20\times 10^{-4}\mpl$, and the future PPTA is expected
to give a bound of $H_*\leq1.41\times 10^{-5}\mpl$.

In the end, let us discuss the most general case with free
parameters $n_t$ and $w$. In this case, the inequality (\ref{e23})
becomes the constraint on the physical parameters $n_t$ and $H_*$
as follows
 \be\label{e24}
 \log_{10}(\frac{H_*}{\mpl})-1.69n_t \le
 \frac{1}{2}(U(\alpha)-6.74\alpha-1.25).
 \ee
Here, we should remember that $\alpha$ relates to the physical
parameters by Eq. (\ref{e22}). The spectra index $n_t$ influences
the bound of $H_*$ mainly by slight changing the corresponding
relation between $\alpha$ and $w$. In Fig. \ref{figure4}, we
calculate the upper bound of $H_*$ in two special cases with
$n_t=0.5$ and $n_t=-0.5$, and compare them with those in the case
of $n_t=0$. This figure shows that the parameter $n_t$ only
slightly affects the bound of $H_*$, and a larger $n_t$ follows a
looser bound of $H_*$. For example, the current PPTA observations
follow $H_*\leq1.82\times10^{-1}\mpl$ at the case with $n_t=0.5$
and $w=0$, which is only $1.6$ times larger than the bound
$H_*\leq1.14\times10^{-1}\mpl$ at the case with $n_t=0$ and $w=0$.

~


{\bf Acknowledgements:} This work is supported by NSFC Grant
Nos.10703005, 10775119 and 11075141.


\end{document}